\documentclass[prd,aps,12pt,showpacs,nofootinbib,tightenlines]{revtex4-1}
\usepackage{mathrsfs}
\usepackage{amsmath}
\usepackage{amssymb}
\usepackage{epsfig}
\usepackage{graphicx}
\usepackage{booktabs}
\usepackage{multirow}
\usepackage{subfigure}
\usepackage{bm}
\usepackage{times}
\usepackage{braket}
\usepackage{color}
\usepackage{slashed}
\usepackage{hyperref}
\newcommand{\beq}{\begin{eqnarray}}
\newcommand{\eeq}{\end{eqnarray}}
\newcommand{\non}{\nonumber\\ }

\definecolor{Red}{rgb}{1.,0.,0.}

\definecolor{Blue}{rgb}{0.,0.,1.}

\definecolor{nicered}{rgb}{0.7,0.1,0.1}
\definecolor{nicegreen}{rgb}{0.1,0.5,0.1}
\hypersetup{colorlinks,citecolor=nicegreen,linkcolor=nicered}

\def \epjc{ Eur. Phys. J. C }

\def \npb{  Nucl. Phys. B }
\def \plb{  Phys. Lett. B }
\def \ppnp{ Prog. Part. Nucl. Phys. }
\def \prd{  Phys. Rev. D }
\def \prl{  Phys. Rev. Lett.  }

\def \jhep{ J. High Energy Phys. }
\begin{document}

\title{$P$-wave contributions to $B_{(s)}\to\psi K\pi$ decays in perturbative QCD approach}

\author{Ya Li$^1$}                \email{liyakelly@163.com}
\author{Zhou Rui$^2$}              \email{jindui1127@126.com}
\author{Zhen-Jun Xiao$^{3}$}    \email{xiaozhenjun@njnu.edu.cn}
\affiliation{$^1$ Department of Physics, College of Sciences, Nanjing Agricultural University, Nanjing, Jiangsu 210095, P.R. China}
\affiliation{$^2$ College of Sciences, North China University of Science and Technology,
                          Tangshan 063009,  P.R. China}
\affiliation{$^3$ Department of Physics and Institute of Theoretical Physics,
                          Nanjing Normal University, Nanjing, Jiangsu 210023, P.R. China}
\date{\today}

\begin{abstract}
In this work, we studied the quasi-two-body decays $B_{(s)} \to \psi [K^*(892), K^*(1410),$ $K^*(1680)] \to \psi K\pi$ by employing the
perturbative QCD (PQCD) factorization approach, where the charmonia $\psi$ represents $J/\psi$ and $\psi(2S)$.
The corresponding decay channels are studied by constructing the kaon-pion distribution amplitude (DA) $\Phi_{K \pi}^{\rm P}$,
which contained the important final state interactions between the kaon and pion in the resonant region.
The relativistic Breit-Wigner formulas are adopted to parameterize the time-like form factor $F_{K\pi}$ appeared in the kaon-pion DAs.
The SU(3) flavor symmetry breaking effect resulting from the mass difference between kaon and pion is taken into account,
which makes significant contributions to the longitudinal polarizations.
We accommodate well the observed branching ratios and the polarization fractions of the $B_{(s)} \to \psi K^*(892) \to \psi K\pi$ by tuning the
hadronic parameters for the kaon-pion DAs.
The PQCD predictions for the $B_{(s)} \to \psi [K^*(1410), K^*(1680)] \to \psi K\pi$ modes from the same set of parameters can be tested
by the future precise data from the LHCb and the Belle II experiments.
\end{abstract}

\maketitle

\section{Introduction}

The $B$ meson decays to heavy vector particles with charmonia and kaon-pion pair, such as $B \to J/\psi K\pi, \psi(2S)K\pi$ ({\it etc}),
have triggered considerable experimental and theoretical attentions in understanding the three-body hadronic $B$ decays.
The strong interest in the polarization and $CP$-asymmetry measurements in $B \to \psi K^*$ decays is motivated by their potential sensitivity
to the new physics beyond the standard model (SM) in the $b \to s$ transition.
In SM, $CP$ violation in $b \to s$ transitions is expected to be very small.
Thus, any significant observation of $CP$ violation may indicate a signal beyond the SM.
Because the precision is still far from the measurements using the tree level processes, this is a new area of research in $B$ physics and there leaves a lot room for contributions from new physics.
In addition, the mixing-induced $CP$-violating asymmetry is measured in the $B^0 \to J/\psi K^{*0}$ decay, where angular analysis allows one to separate $CP$-eigenstate amplitudes.
This allows one to resolve the sign ambiguity of the $\cos 2\beta$ term that appears in the time-dependent angular distribution due to interference of parity-even and parity-odd terms. 
The detailed amplitude analyses of the $B \to \psi K\pi$ decays have been performed by the BABAR~\cite{prd71-032005,prl94-141801,prd76-031102},
Belle~\cite{plb538-11,prl95-091601,prd80-031104,prd88-072004,prd90-112009}, LHCb~\cite{epjc72-2118,prd88-052002,jhep11-082,plb747-484},
CDF~\cite{prl76-2015,prd58-072001,prl85-4668,prl94-101803,prd83-052012}, CLEO~\cite{prl79-4533} and D0 Collaboration~\cite{prl102-032001}.

On the theory side, the three-body $B$ meson decays do receive both the resonant and nonresonant contributions, as well as the
possible significant final-state interactions (FSIs)~\cite{prd89-094013,1512-09284,prd89-053015}.
Since the nonresonant contributions and possible FSIs can not be evaluated reliably,
the analysis for the three-body hadronic decays of the $B$ meson are much more complicated than those for the two-body decays.
Fortunately, the validity of factorization for these kinds of $B$ decays can be assumed in the quasi-two-body mechanism where all possible interactions between the pair of the mesons are included but the interactions between the bachelor particle and  the daughter mesons from the resonance are ignored.
Several theoretical approaches for describing three-body  hadronic decays of $B$ meson based on the symmetry principles and factorization theorems have been developed.
The QCD-improved factorization (QCDF)~\cite{prl83-1914,npb591-313,npb606-245,npb675-333} has been widely applied in the study of the three-body charmless hadronic decays of $B$ meson~\cite{npb899-247,plb622-207,prd74-114009,prd79-094005,APPB42-2013,prd76-094006,prd88-114014,prd94-094015,prd89-094007,prd87-076007,jhep10-117}.
The $U$-spin and flavor $SU(3)$ symmetries were also adopted to analyse the three-body decays in Refs.~\cite{prd72-094031,plb727-136,prd72-075013,prd84-056002,plb728-579,prd91-014029}.

As is well known, there are several theoretical approaches that can be used to calculate the hadronic $B$ meson decays, such as the QCD-improved factorization (QCDF)~\cite{prl83-1914,npb591-313,npb606-245,npb675-333}, the perturbative QCD (PQCD) factorization approach~\cite{prd63-074009,plb504-6,ppnp51-85} and the soft-collinear-effective theory (SCET)~\cite{prd70-054015,prd74-034010,npb692-232,prd72-098501,prd72-098502}.
For most $B \to h_1 h_2$ decay channels, the theoretical predictions obtained by adopting these different
factorization approaches agree well with each other and also are well consistent with the data within errors.
It has been known that the naive factorization assumption (FA) does not apply to exclusive $B$ meson decays into charmonia, such as $B \to J/\psi K$~\cite{prd59-092004}, which belongs to the color-suppressed mode.
For a color-suppressed mode, a significant impact of nonfactorizable contribution is expected.
The predictions from FA for such decay channels are always small since they neglect the nonfactorizable effects.
Thus, many attempts to resolve this puzzle have been made in more sophisticated approaches (for a review, see~\cite{ppnp51-85}).
Although the nonfactorizable corrections to the FA have been included in QCDF approach, their predictions are still too small to explain the data.
As we know, QCDF is based on the collinear factorization, in which $B$ meson transition form factors suffer the end-point singularity.
The end-point singularity may render the estimation of the nonfactorizable contributions out of control.
More details can be seen in Ref.~\cite{prd71-114008}.
Based on the $k_T$ factorization theorem, the perturbative QCD (PQCD) approach~\cite{prl74-4388,plb348-597} is suitable for describing different types of heavy hadron decays.
The transverse-momentum-dependent hadronic wave function is introduced to remove the potential light-cone divergence and the rapidity singularity~\cite{JHEP06-013,JHEP02-008}.
Then both factorizable and nonfactorizable contributions are calculable without end-point singularity.
The Sudakov resummation has also been introduced to suppress the long-distance contributions effectively.
Therefore, the PQCD approach is a self-consistent framework and has a good predictive power.
In the previous works~\cite{prd71-114008,prd89-094010,prd90-114030,epjc77-610,epjc75-293,epjc60-107},
the two-body decays of the $B (B_c)$ mesons to $\psi$ plus a light vector meson have been  studied in the PQCD framework,
while the PQCD  predictions are in good agreement with the  data.
However, the width of the resonant state and the interactions between the final states associated with the resonances play an important role on the branching ratios and the direct $CP$ violations of the quasi-two-body decays in
Refs.~\cite{plb763-29,prd95-056008,prd96-093011,prd97-033006,prd98-056019,prd97-034033,epjc79-37,plb791-342,epjc79-539}.
It seems more appropriate to treat a light vector meson as an intermediate resonance.

\begin{figure}[tbp]
\centerline{\epsfxsize=14cm \epsffile{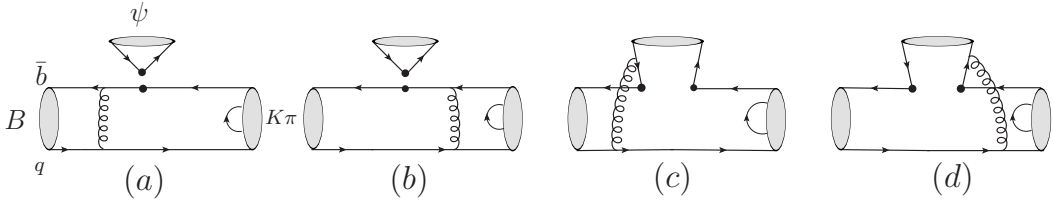}}
\caption{Typical leading-order Feynman diagrams for the quasi-two-body decays $B \to \psi (K^* \to) K \pi$,
with $q=(u,d,s)$. Diagrams (a) and (b) represent the factorizable contributions, and diagrams (c) and (d) denote the
nonfactorizable contributions.}
\label{fig:fig1}
\end{figure}

In this paper,  with the help of the $P$-wave kaon-pion DAs along with the time-like form factor $F_{K\pi}$ which contains the final-state interactions
between kaon-pion pair, the quasi-two-body decays $B\to \psi [K^*(892),K^*(1410), K^*(1680)]\to \psi K\pi$, as shown in Fig.~\ref{fig:fig1},
will be studied in the PQCD approach utilizing framework discussed in~\cite{plb561-258,prd70-054006},
albeit the underlying $k_T$ factorization has not been proven rigorously ~\cite{prd88-114014,prd94-094015}.
Throughout the remainder of the paper, the symbol $K^*$ is used to denote the $K^*(892)$ resonance.
As the spin of $\psi$ meson is 1, there are three possible polarizations generating the longitudinal (0), parallel ($\parallel$), and perpendicular ($\perp$)
amplitudes.
Therefore, the $K\pi$ DAs involving both longitudinal and transverse polarizations are nontrivial nonperturbative inputs in our calculations.
The two-pion (two-kaon) DAs corresponding to both longitudinal and transverse polarizations have been constructed to capture
important final state interactions in the processes involving the resonant $\rho$ ($\phi$) in our previous works~\cite{prd98-113003,1907-04128}.
The $P$-wave kaon-pion DAs are introduced similar to the case of two-pion ones~\cite{prd98-113003} and the SU(3) flavor
symmetry breaking effect for the kaon-pion pair is considered, which plays an important role in the longitudinal polarizations.

For the considered three-body hadronic $B$ meson decays,  the leading contributions are identified by defining power counting rules for
various topologies of amplitudes~\cite{plb561-258}.
There is no proof of factorization for the three-body $B$ decays at present.
We can only restrict ourselves to specific kinematical configurations on which our work is based.
The Dalitz plot contains different regions with ``specifical'' kinematics.
The central region corresponds to the case where all three final particles fly apart with large energy
and none of them moves collinearly to any other.
This situation called a genuine three-body decay, which contains two hard gluons is power and $\alpha_s$ suppressed compared to the leading contribution.
The corners correspond to the case in which one final particle is
approximately at rest (i.e. soft), and the other two are fast and  back-to-back.
The central part of the edges corresponds to the case in which two particles move collinearly and the other particle recoils back.
More details can be referred to Refs.~\cite{npb899-247,1609-07430}.
The significance of these special kinematic configurations is that different theoretical approaches may be applicable in these different regions.
The specific kinematic configuration, where two particles are collinear and generate a small invariant mass recoiling against the third one, gives a dominant contribution.
This situation exists particularly in the low $\pi\pi$ or $K\pi$ invariant mass region ($\lesssim$2 GeV) of the Dalitz plot.
It seems appropriate to assume the validity of factorization for this quasi-two-body $B$ meson decay.
Then it is reasonable to assume that the dynamics associated with the pair of final state mesons can be factorized into a two-meson distribution amplitude $\Phi_{h_1h_2}$~\cite{MP,MT01,MT02,MT03,NPB555-231,Grozin01,Grozin02}.
The contribution from the soft kaon or pion region has been included into the two-meson DA, because it also corresponds to the region with a small invariant mass.

The typical PQCD factorization formula for a $B\to \psi K\pi$ decay amplitude is written in the following form~\cite{plb561-258},
\begin{eqnarray}
\mathcal{A}=\Phi_B\otimes H\otimes \Phi_{K\pi}\otimes\Phi_{\psi}.
\end{eqnarray}
The hard kernel $H$ includes the dynamics of the strong and electroweak interactions in three-body hadronic decays in a similar way as the one for the corresponding two-body decays.
The $\Phi_B$ ($\Phi_{\psi}$) denotes the wave function for the $B$ meson (the final state meson $\psi$).
The $\Phi_{K\pi}$ is the two-hadron distribution
amplitude, which absorbs the nonperturbative dynamics in the $K$-$\pi$  hadronization process.

The layout of the present paper is as follows.
In Sec.~II, we briefly introduce the theoretical framework.
The numerical values and some discussions will be given in Sec.~III.
Section IV contains our conclusions.
The Appendix collects the explicit PQCD factorization formulas for all the decay amplitudes.

\section{FRAMEWORK}\label{sec:2}
We will work in the $B$ meson rest frame and employ the light-cone coordinates for momentum variables.
The $B$ meson momentum $p_{B}$, the total momentum of the kaon-pion pair,
$p=p_1+p_2$, the final-state $\psi$ momentum $p_3$  and the quark momentum $k_i$ in each meson are chosen as
\begin{eqnarray}
p_{B}&=&\frac{m_{B}}{\sqrt 2}(1,1,0_{\rm T}),~\quad p=\frac{m_{B}}{\sqrt2}(1-r^2,\eta,0_{\rm T}),~\quad
p_3=\frac{m_{B}}{\sqrt 2}(r^2,1-\eta,0_{\rm T}), \nonumber\\
k_{B}&=&\left(0,x_B \frac{m_{B}}{\sqrt2} ,k_{B \rm T}\right),\quad
k= \left( z (1-r^2)\frac{m_{B}}{\sqrt2},0,k_{\rm T}\right),\non
k_3&=&\left(r^2x_3\frac{m_B}{\sqrt{2}},(1-\eta)x_3 \frac{m_B}{\sqrt{2}},k_{3{\rm T}}\right),\label{mom-B-k}
\end{eqnarray}
where $m_{B}$ is the mass of $B$ meson, $\eta=\frac{\omega^2}{(1-r^2)m^2_{B}}$ with $r=m_{\psi}/m_{B}$, $m_{\psi}$ is the
mass of charmonia, and the invariant mass squared $\omega^2=p^2$. The momentum fractions $x_{B}$, $z$ and $x_3$ run from zero to unity.

As usual we also define the momentum $p_1$ and $p_2$ of kaon-pion pair as
\begin{eqnarray}\label{eq:p1p2}
 p_1&=&\left (\zeta p^+, (1-\zeta)\eta p^+, \sqrt{\zeta(1-\zeta)}\omega,p_{1\rm T} \right ),\non
  p_2&=& \left ( (1-\zeta) p^+, \zeta\eta p^+, -\sqrt{\zeta(1-\zeta)}\omega,p_{2\rm T} \right ),
\end{eqnarray}
with $\zeta=p_1^+/P^+$ characterizing the distribution of the longitudinal momentum of kaon.

The kaon-pion DAs can be related to the single kaon and pion through a perturbative
evaluation of the matrix elements \cite{prd70-054006,prd97-034033},
 \begin{eqnarray}\label{eq:matrix}
\langle K(p_1)\pi(p_2)|\bar{q}_1(y^-)\Gamma q_2(0)|0\rangle,
\end{eqnarray}
as a time-like kaon-pion production process, where $\Gamma$ denotes the possible spin projectors $I$, $\gamma_5$,
$\gamma_{\mu}$, $\gamma_{\mu}\gamma_5$, $\sigma_{\mu\nu}$, and $\sigma_{\mu\nu}\gamma_5$.
All these  projectors, except for the vector one, are evaluated in the same way as those for the two-pion ones~\cite{prd98-113003},
while the vector current matrix elements should be reanalysed in this work
since the significant SU(3) flavor symmetry breaking effect  are included as aforementioned.

The matrix element for the transition from vacuum to the $K\pi$ state via the vector current is defined in terms of the vector $F^{\parallel}_{K\pi}(s)$ and scalar $F^0_{K\pi}(s)$ form factors as follows,
\begin{eqnarray}
\langle K(p_1)\pi(p_2)|\bar{q}_1\gamma_{\mu}q_2|0\rangle&=&\left[(p_1-p_2)_{\mu}-\frac{m^2_{K}-m^2_{\pi}}{(p_1+p_2)^2}(p_1+p_2)_{\mu}\right]
F^{\parallel}_{K\pi}(s)\non
&& +\frac{m^2_K-m^2_{\pi}}{(p_1+p_2)^2}(p_1+p_2)_{\mu}F^0_{K\pi}(s),
\end{eqnarray}
with the invariant mass squared $s=\omega^2=(p_1+p_2)^2$.
The scalar strange resonances contribute to the scalar form factor $F^0_{K\pi}(s)$, while the vector resonances contribute to the vector form factor $F^{\parallel}_{K\pi}(s)$.
This work focuses on the vector strange resonances $K^*(892),K^*(1410),K^*(1680)$.
Note that the second term in the bracket appears because of the unequal mass between $K$ and $\pi$.
The term $(m^2_{K}-m^2_{\pi})/(p_1+p_2)^2$ describes the SU(3) symmetry breaking effect.

One can apply the parametrization for the longitudinal and transverse components corresponding to the $\gamma_{\mu}$ spin projector,
\begin{itemize}
\item[(1)]
 For the case of $\mu=+$, we get   $(p_1-p_2)_\mu = (2\zeta-1 ) p^+$, $(p_1+p_2)_\mu =p^+$, the twist-2 DA of longitudinal polarization can be described as
\beq
{p \hspace{-1.5truemm}/} \left [P_1(2\zeta-1)-\frac{m^2_{K}-m^2_{\pi}}{\omega^2}P_0(2\zeta-1)\right ]\phi^0,
\eeq
where $P_0(2\zeta-1)=1$ and $P_1(2\zeta-1)=2\zeta-1$ are two Legendre Polynomials.
\item[(2)]
 For the case of $\mu=x$, we get
 \beq
 (p_1-p_2)^x = 2 p_1^x= -2\sqrt{\zeta(1-\zeta)}\omega
                       \frac{\epsilon^{x\nu\rho\sigma}
                       \epsilon_{T\nu} p_{\rho} n_{-\sigma}}{p\cdot n_-}, \quad (p_1+p_2)^x=0,
\eeq
where the transverse polarization vector is normalized into~\cite{prd98-113003}
\begin{eqnarray}\label{eq:jihua}
\epsilon_{T\mu} = \frac{\epsilon_{\mu\nu\rho\sigma}
                  p_1^{\nu} p^{\rho} n_{-}^{\sigma}}{\sqrt{\zeta(1-\zeta)}\; \omega p\cdot n_-}.
\end{eqnarray}
\end{itemize}
Obviously, the SU(3) asymmetry term  contributes to the longitudinal twist-2 DA but not the transverse one.

Following the prescription in Refs.~\cite{prd98-113003,plb730-336}, the expansions of the nonlocal matrix elements for various spin projectors $\Gamma$ up to twist-3 are listed below:
\begin{eqnarray}\label{eq:v}
\langle K(p_1)\pi(p_2) |\bar{q}_1(y^-)\gamma_{\mu}q_2(0)|0\rangle&=&(2\zeta-1-\frac{m^2_{K}-m^2_{\pi}}{\omega^2})p_{\mu}\int_0^1
dze^{izP\cdot y}\phi^0(z,\omega)\non
&-&2\sqrt{\zeta(1-\zeta)}\omega \frac{\epsilon^{\mu\nu\rho\sigma}\epsilon_{T\nu}p_{\rho}n_{-\sigma}}{p\cdot n_-}
\int_0^1 dze^{izp\cdot y}\phi^v(z,\omega),
\end{eqnarray}
\begin{eqnarray}\label{eq:s}
\langle K(p_1)\pi(p_2)|\bar{q}_1(y^-)Iq_2(0)|0\rangle&=&\omega \int_0^1
dze^{izp\cdot y}\phi^s(z,\omega),
\end{eqnarray}
\begin{eqnarray}\label{eq:t}
\langle K(p_1)\pi(p_2) |\bar{q}_1(y^-)\sigma_{\mu\nu}q_2(0)|0\rangle&=&-i
\frac{p_{1\mu}p_{2\nu}-p_{1\nu}p_{2\mu}}{\omega } \int_0^1
dze^{izp\cdot y}\phi^t(z,\omega),
\end{eqnarray}
\begin{eqnarray}\label{eq:ta}
\langle K(p_1)\pi(p_2) |\bar{q}_1(y^-)\sigma_{\mu\nu}\gamma_5q_2(0)|0\rangle&=&-\sqrt{\zeta(1-\zeta)} \epsilon_{T\nu}p_{\mu} \int_0^1
dze^{izp\cdot y}\phi^T(z,\omega),
\end{eqnarray}
\begin{eqnarray}\label{eq:va}
\langle K(p_1)\pi(p_2) |\bar{q}_1(y^-)\gamma_{\mu}\gamma_5q_2(0)|0\rangle&=& i\sqrt{\zeta(1-\zeta)}\omega \epsilon_{T\mu} \int_0^1
dze^{izp\cdot y}\phi^a(z,\omega),
\end{eqnarray}
\begin{eqnarray}
\langle K(p_1)\pi(p_2) |\bar{q}_1(y^-)\gamma_5q_2(0)|0\rangle&=&0,
\end{eqnarray}
with the kaon-pion DAs $\phi^{0,T}$ and $\phi^{s,t,v,a}$ being of twsit-2 and twist-3, respectively.
It is shown that the SU(3) asymmetry factor $(m^2_{K}-m^2_{\pi})/\omega^2$ only exist in the longitudinal DA $\phi^0$, but not in the other DAs up to twist 3.

The $P$-wave kaon-pion DAs related to both longitudinal and transverse polarizations are introduced in analogy with the case of two-pion ones~\cite{prd98-113003},
\begin{eqnarray}
\Phi_{K\pi}^{L}&=&\frac{1}{\sqrt{2N_c}} \left [{ p \hspace{-1.5truemm}/ }\phi^0(z,\zeta,\omega^2)+\omega\phi^s(z,\zeta,\omega^2)
+\frac{{p\hspace{-1.5truemm}/}_1{p\hspace{-1.5truemm}/}_2
  -{p\hspace{-1.5truemm}/}_2{p\hspace{-1.5truemm}/}_1}{\omega(2\zeta-1)}\phi^t(z,\zeta,\omega^2) \right ] \;,\non
\Phi_{K\pi}^{T}&=&\frac{1}{\sqrt{2N_c}}
\Big [\gamma_5{\epsilon\hspace{-1.5truemm}/}_{T}{ p \hspace{-1.5truemm}/ } \phi^T(z,\zeta,\omega^2)
+\omega \gamma_5{\epsilon\hspace{-1.5truemm}/}_{T} \phi^a(z,\zeta,\omega^2)\non
&& + i\omega\frac{\epsilon^{\mu\nu\rho\sigma}\gamma_{\mu}
\epsilon_{T\nu}P_{\rho}n_{-\sigma}}{P\cdot n_-} \phi^v(z,\zeta,\omega^2) \Big ].
\label{eq:phifunc}
\end{eqnarray}
The key in our work is to get the expressions of $\phi^0(z,\zeta,\omega^2)$.
According to Eq.~(2.9) in Ref.~\cite{NPB555-231}, we decompose the kaon-pion DA in eigenfunctions of the evolution equation (Gegenbauer polynomials $C_n^{\frac{3}{2}}(2z-1)$).
Eventually, the explicit expression of $\phi^0(z,\zeta,\omega^2)$ can be written in the following form:
\begin{eqnarray}
\phi^0(z,\zeta,\omega^2)&=&\frac{F^{\parallel}_{K\pi}}{2\sqrt{2}}6z(1-z)\left[\sum a_nC_n^{\frac{3}{2}}(2z-1)\right](2\zeta-1-\alpha).
\end{eqnarray}
Below we label the SU(3) symmetry breaking effect $(m^2_{K}-m^2_{\pi})/\omega^2$ by $\alpha$ for simplicity.

The various twist DAs $\phi^i$ have similar forms as the corresponding ones for the $K^*$ meson~\cite{prd71-114008} by replacing the
decay constants with the time-like form factor,
\begin{eqnarray}
\phi^0(z,\zeta,\omega^2)&=&\frac{3F_{K\pi}^{\parallel}(\omega^2)}{\sqrt{2N_c}} z(1-z)\left[1+a_{1K^*}^{||}3t+a_{2K^*}^{||}\frac{3}{2}(5t^2-1)\right](2\zeta-1-\alpha)\;,\label{eqphi0}\\
\phi^s(z,\zeta,\omega^2)&=&\frac{3F_{K\pi}^{\perp}(\omega^2)}{2\sqrt{2N_c}}\bigg\{ t\left[1+a_{1s}^{\perp} t\right]-a_{1s}^{\perp}2z(1-z)\bigg\}(2\zeta-1) \;,\label{eqphis}\\
\phi^t(z,\zeta,\omega^2)&=&\frac{3F_{K\pi}^{\perp}(\omega^2)}{2\sqrt{2N_c}} t\left[ t+a_{1t}^{\perp}(3 t^2-1)\right](2\zeta-1) \;,\label{eqphit}\\
\phi^T(z,\zeta,\omega^2)&=&\frac{3F_{K\pi}^{\perp}(\omega^2)}{\sqrt{2N_c}}z(1-z)\left[1+a_{1K^*}^{\perp}3 t+
a_{2K^*}^{\perp}\frac{3}{2}(5 t^2-1)\right]\sqrt{\zeta(1-\zeta)}\;,\label{eqphi}\\
\phi^a(z,\zeta,\omega^2)&=&\frac{3F_{K\pi}^{\parallel}(\omega^2)}{4\sqrt{2N_c}}\bigg\{ t\left[1+a_{1a}^{\parallel} t\right]-a_{1a}^{\parallel}2z(1-z)\bigg\}\sqrt{\zeta(1-\zeta)}\;,\label{eqphia}\\
\phi^v(z,\zeta,\omega^2)&=&\frac{3F_{K\pi}^{\parallel}(\omega^2)}{8\sqrt{2N_c}}\left[1+ t^2+a_{1v}^{\parallel} t^3\right]\sqrt{\zeta(1-\zeta)}\;,\label{eqphiv}
\end{eqnarray}
where $t=(1-2z)$, and we introduce two Gegenbauer moments $a_1$ and $a_2$ for the twist-2 DAs and one Gegenbauer moment $a_1$ for each twist-3 one.
While the $B$ meson and $\psi$ DAs are the same as those widely adopted in the PQCD approach~\cite{prd90-114030,epjc75-293,epjc77-610,epjc60-107,prd98-113003}.

The relativistic Breit-Wigner (RBW) line shape is adopted for the $P$-wave resonance $K^*(892),$
$ K^*(1410)$ and $ K^*(1680)$ to parameterize the time-like form factor $F_{K\pi}^{\parallel}(s)$, which is widely adopted in the experimental data analysis.
The explicit expression is in the following form~\cite{prd83-112010},
\begin{eqnarray}
F_{K\pi}^{\parallel}(s)&=&\frac{c_1 m_{K^*(892)}^2}{m^2_{K^*(892)} -s-im_{K^*(892)}\Gamma_1(s)}
+\frac{c_2 m_{K^*(1410)}^2}{m^2_{K^*(1410)} -s-im_{K^*(1410)}\Gamma_2(s)}\non
&& + \frac{c_3 m_{K^*(1680)}^2}{m^2_{K^*(1680)} -s-im_{K^*(1680)}\Gamma_3(s)}\;,
\end{eqnarray}
where $c_i$ (i=1,2,3) are the corresponding weight coefficients and satisfy $c_1+c_2+c_3=1$ due to the normalization condition $F_{K\pi}^{\parallel}(0)=1$.
The three terms describe the contributions from $K^*(892)$, $K^*(1410)$, and $K^*(1680)$, respectively.
We find that there is no existing data for the interferences among these resonances.
Thus, as a first order approximation, we only determine the modules of the complex weight coefficients $c_i$ (i=1,2,3) and ignore their phases.

Here, the mass-dependent width $\Gamma_i(s)$ is defined by
\begin{eqnarray}
\Gamma_i(s)&=&\Gamma_i\frac{m_i}{\sqrt{s}}\left(\frac{|\overrightarrow{p_1}|}{|\overrightarrow{p_0}|}\right)^{(2L_R+1)},
\end{eqnarray}
where $|\overrightarrow{p_1}|$ is the momentum vector of the resonance decay product measured in the resonance rest frame, and $|\overrightarrow{p_0}|$ is the value of $|\overrightarrow{p_1}|$ when $\sqrt{s}=m_{K^*}$.
$L_R$ is the orbital angular momentum in the $K\pi$ system and $L_R=0,1,2,...$ corresponds to the $S,P,D,...$ partial-wave resonances.
Since the vector resonance has spin-1 and kaon (pion) with spin-0, when the $K\pi$ system forms a spin-1 resonance, the orbital angular momentum between the kaon and pion must be $L_R=1$, which refers to a $P$ wave configuration.
The $m_i$ and $\Gamma_i$ are the pole mass and width of the corresponding resonance, where $i=1,2,3$ represents the resonance
$K^*(892)$, $K^*(1410)$ and $K^*(1680)$, respectively.
Following Ref.~\cite{plb763-29}, we also assume that
\begin{eqnarray}
F_{K\pi}^{\perp}(s)/F_{K\pi}^{\parallel}(s)\approx f_{K^*}^T/f_{K^*}.
\end{eqnarray}
with $f_{K^*}=0.217 \pm 0.005 {\rm GeV}, f^T_{K^*}=0.185 \pm 0.010 {\rm GeV}$~\cite{prd70-034009,prd76-074018,prd85-094003}.
Due to the limited studies on the decay constants of $K^*(1410)$ and $K^*(1680)$, we use the two decay
constants of $K^*(892)$ to determine the ratio $F_{K\pi}^{\perp}(s)/F_{K\pi}^{\parallel}(s)$.

\section{Numerical results}\label{sec:3}

The differential branching fraction for the $B\rightarrow \psi K\pi$ decays into
$P$-wave kaon-pion pair is expressed as
\begin{eqnarray}\label{eq:dfenzhibi}
\frac{d \mathcal{B}}{d \omega}=\frac{\tau_B \omega|\vec{p}_1||\vec{p}_3|}{32\pi^3M^3}
\sum_{i=0,\parallel,\perp}|\mathcal{A}_i|^2,
\end{eqnarray}
where the kaon and charmonium three-momenta in the $K\pi$ center-of-mass frame are given by
\begin{eqnarray}
|\overrightarrow{p_1}|=\frac{\sqrt{\lambda(\omega^2,m_K^2,m_{\pi}^2)}}{2\omega}, ~\quad
|\overrightarrow{p_3}|=\frac{\sqrt{\lambda(m_B^2,m_{\psi}^2,\omega^2)}}{2\omega},
\end{eqnarray}
with the kaon (pion) mass $m_K$ ($m_{\pi})$ and the K$\ddot{a}$ll$\acute{e}$n function
$\lambda (a,b,c)= a^2+b^2+c^2-2(ab+ac+bc)$.
The terms $\mathcal{A}_0$, $\mathcal{A}_{\parallel}$ and $\mathcal{A}_{\perp}$ represent
the longitudinal, parallel and perpendicular polarization amplitudes in the transversity
basis, respectively, which are related to $\mathcal{A}_{L,N,T}$ in the Appendix.
The polarization fractions $f_{\lambda}$ with $\lambda=0$, $\parallel$,
and $\perp$ are described as
\begin{eqnarray}\label{pol}
f_{\lambda}=\frac{|\mathcal{A}_{\lambda}|^2}{|\mathcal{A}_0|^2
+|\mathcal{A}_{\parallel}|^2+|\mathcal{A}_{\perp}|^2},
\end{eqnarray}
with the normalisation relation $f_0+f_{\parallel}+f_{\perp}=1$.

Before proceeding with the numerical analysis, the meson masses and widths (in units of GeV) are collected below for the numerical calculations~\cite{pdg2018}:
\begin{eqnarray}
m_{B}&=&5.280, \quad m_{B_s}=5.367, \quad m_{K^{*0}}=0.89555, \quad m_{K^*(1410)}=1.421,\quad m_{K^*(1680)}= 1.718, \nonumber\\
m_{J/\psi}&=&3.097, \quad m_{\psi(2S)}=3.686, \quad m_{\pi^{\pm}}=0.140, \quad m_{K^{\pm}}=0.494, \nonumber\\
\quad \Gamma_{K^*}&=&0.0473, \quad \Gamma_{K^*(1410)}=0.236,\quad \Gamma_{K^*(1680)}=0.322.
\end{eqnarray}
The values of the Wolfenstein parameters are adopted as given in the Ref.~\cite{pdg2018}:
$A=0.836\pm0.015, \lambda=0.22453\pm 0.00044$, $\bar{\rho} = 0.122^{+0.018}_{-0.017}$, $\bar{\eta}= 0.355^{+0.012}_{-0.011}$.
The decay constants (in units of GeV) and the $B$ meson lifetimes (in units of ps) are chosen
as~\cite{prd95-056008,prd90-114030,epjc75-293}
\begin{eqnarray}
f_B&=&0.19, \quad f_{B_s}=0.23, \quad  f_{J/\psi}=0.405, \quad  f_{\psi(2S)}=0.296, \quad f_{K^*}=0.217,
 \quad f^T_{K^*}=0.185,\nonumber\\
\tau_{B^0}&=&1.519,\quad \tau_{B^{\pm}}=1.638, \quad \tau_{B_{s}}=1.512.
\end{eqnarray}

\begin{figure}[tbp]
\centerline{\epsfxsize=10cm \epsffile{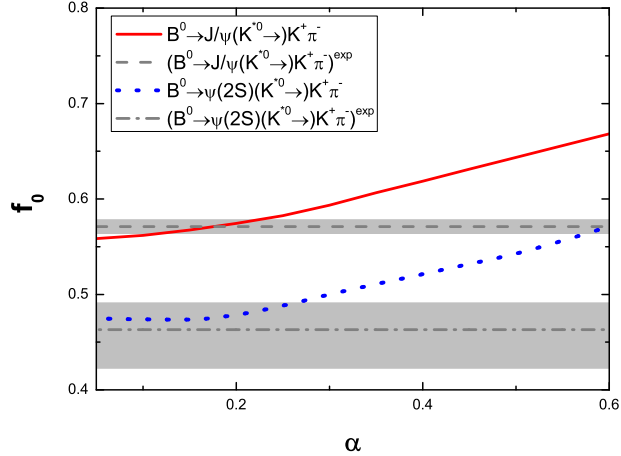}}
\caption{The longitudinal polarization fraction $f_0$ as a function of the asymmetric factor $\alpha$ for the $B^0 \to J/\psi (K^{*}(892)^{0} \to)K^+\pi^- $ (red solid line) and $B^0 \to \psi(2S) (K^{*}(892)^{0} \to)K^+\pi^- $ (blue dotted line) decays . The gray dashed and dash-dotted lines represent the central value of the experimental data~\cite{pdg2018} for the corresponding decay modes, respectively. Shaded bands show the experimental errors.}
\label{fig:fig2}
\end{figure}

The Gegenbauer moments and coefficients $c_i$ (i=1,2,3) are determined on basis of the existing data for the $B\to \psi K\pi$ branching ratios and polarization fractions from PDG2018~\cite{pdg2018},
\begin{eqnarray}
a_{1K^*}^{\parallel}&=&0.2, ~\quad a_{2K^*}^{\parallel}=0.5,~\quad a_{1s}^{\perp}=-0.2,~\quad a_{1t}^{\perp}=0.2,\nonumber\\
a_{1K^*}^{\perp}&=&0.3, ~\quad a_{2K^*}^{\perp}=0.8, ~\quad a_{1a}^{\parallel}=-0.3,~\quad a_{1v}^{\parallel}=0.3,\nonumber\\
c_1&=&0.72, ~\quad c_2=0.135, ~\quad c_3=0.145.
\end{eqnarray}
It should be stressed that, unlike the asymptotic forms of the twist-3 DAs used in the two-body work~\cite{prd89-094010}, we introduce one Gegenbauer moment $a_1$ for each twist-3 DA in analogous to the cases for the resonance $\rho$ in Ref.~\cite{prd98-113003}.
Meanwhile, we consider the important asymmetric factor $\alpha$, which has strong influence on the Gegenbauer moments of twist-2 DAs.
The combined effect makes the Gegenbauer moments of twist-2 DAs in the quasi-two-body framework different from those for two-body ones~\cite{prd89-094010}.
In this work, we fix the Gegenbauer moments of the kaon-pion DAs by matching the theoretical results to the experimental data.
To determine the four longitudinal Gegenbauer moments $a_{1K^*}^{\parallel}$, $a_{2K^*}^{\parallel}$, $a_{1s}^{\perp}$, $a_{1t}^{\perp}$, we pick up the four branching ratios associated with the longitudinal
polarization of the $B^0 \to J/\psi (K^{*}(892)^{0} \to)K^+\pi^- $ and $B^0 \to \psi(2S) (K^{*}(892)^{0} \to)K^+\pi^- $ decays from~\cite{pdg2018}, and of the $B^0_s \to J/\psi (\bar{K}^{*}(892)^{0} \to)K^-\pi^+ $ decay measured by LHCb~\cite{jhep11-082}, as well as the branching ratio of the $B^0 \to \eta_c(1S) (K^{*}(892)^{0} \to)K^+\pi^- $ decay~\cite{pdg2018} as the data inputs.
Thus we can solve the four longitudinal Gegenbauer moments from the four inputs.
Similarly, the four transverse ones can be constrained by the transverse polarization fractions of the two decay modes $B^0 \to J/\psi (K^{*}(892)^{0} \to)K^+\pi^- $ and $B^0 \to \psi(2S) (K^{*}(892)^{0} \to)K^+\pi^- $.
Each decay channel has the parallel and perpendicular components.
Then we solve the four transverse Gegenbauer moments from the four inputs as well.
Of course, considering the experimental uncertainties, it is difficult to restrict these parameters precisely.
We just make the theoretical branching ratios compatible with the experimental ones by adjusting the corresponding Gegenbauer moments.
Apart from the results for the $B^0 \to J/\psi (K^{*}(892)^{0} \to) K^+\pi^-$ and $B^0 \to \psi(2S) (K^{*}(892)^{0} \to) K^+\pi^-$ decays from~\cite{pdg2018} and the branching ratio of the $B^0_s \to J/\psi (\bar{K}^{*}(892)^{0} \to)K^-\pi^+ $ decay associated with the longitudinal polarization~\cite{jhep11-082},
other results shown in Table~\ref{tab:br} and \ref{tab:br2} are our predictions based on the determined parameters.

The SU(3) asymmetric term $\alpha=(m^2_{K}-m^2_{\pi})/\omega^2$ plays an important role in the longitudinal polarization fraction $f_0$ since it appears in the expression of the longitudinal twist-2 kaon-pion DA.
We also check the average value of the asymmetric factor $\alpha$.
Its value is estimated to lie in the range from 0.05 to 0.6 by the relation $\alpha=(m^2_{K}-m^2_{\pi})/\omega^2$ according to
the kinematic bounds on the value of $\omega$ [$(m_K+m_\pi) \leq \omega\leq (m_B-m_\psi )$].
We find that the average value of $\alpha$ is about 0.2 effectively, shown in Fig.~\ref{fig:fig2}, where the curves for the central values of the $B^0 \to J/\psi (K^{*}(892)^{0} \to)K^+\pi^- $ (the gray dashed line) and the $B^0 \to \psi(2S) (K^{*}(892)^{0} \to)K^+\pi^- $ (the gray dash-dotted line) decays associated with longitudinal polarization fractions from data~\cite{pdg2018}, and the curves for the $B^0 \to J/\psi (K^{*}(892)^{0} \to)K^+\pi^- $ (the red solid line) and the $B^0 \to \psi(2S) (K^{*}(892)^{0} \to)K^+\pi^- $ (the blue dotted line) decays are obtained by varying $\alpha$ as a free parameter.

Since the $K^*(892)$ components in both $J/\psi$ and $\psi(2S)$ modes are well measured with a high significance
by the Belle Collaboration \cite{prd80-031104,prd90-112009},
we can exactly determine its weight  coefficient $c_1=0.72$ based on its fit fraction.
However, the significance of the two high-mass $K^*$ states are too low for us to determine the $c_2$ and $c_3$ precisely.
For example, the fit fractions for the $K^*(1410)$ and $K^*(1680)$ components in the $\psi(2S)$ channel are $5.5^{+8.8}_{-1.5}\%$
(statistical error only) and $2.8^{+5.8}_{-1.0}\%$ (statistical error only) respectively.
The corresponding measurements in the $J/\psi$ mode for the two components are both $0.3^{+0.2}_{-0.1}\%$ (statistical error only).
We note that LHCb has made the first Dalitz plot analysis for $B^0\rightarrow \eta_c(1S) K^+\pi^-$ decays ~\cite{epjc78-1019} .
Their fitting  model gives almost equal fit fractions of $K^*(1410)$ and $K^*(1680)$ (see Table~7 of Ref.~\cite{epjc78-1019}),
which is very similar with the situation  in $J/\psi$ mode as mentioned above.
Therefore, it is reasonable to assume that the branching ratios of the $B \to \psi K^*(1410)^0 \to \psi K^+\pi^-$ and $B \to \psi K^*(1680)^0\to \psi K^+\pi^-$
are approximately equal and employ the normalization of  $c_1+c_2+c_3=1$, then $c_2$ and $c_3$ are determined as $0.135$ and $0.145$, respectively.
The small gap between them is understandable with respect to the different nominal masses and widths of $K^*(1410)$ and $K^*(1680)$.
We emphasize that our predictions on the excited state channels are only rough estimates of magnitude,
which need to be tested precisely in the future experiments.

By using the Eqs.~(\ref{eq:dfenzhibi}-\ref{pol}), the decay amplitudes in the Appendix and all the input quantities,
the resultant branching ratios $\mathcal{B}$ and the polarization fractions $f_{\lambda}$ together with the available
experimental measurements  are summarized in Table~\ref{tab:br}, while those for $\psi(2S)$ are listed in Table~\ref{tab:br2}.
Because the charged $B$ meson decays differ from the neutral ones only in the lifetimes and the isospin factor in our formalism,
we can derive the branching ratios for the $B^+$ meson via multiplying those for the $B^0$ meson by the ratio $\tau_{B^+}/\tau_{B^0}$.
For the color-suppressed decays, it is expected that the factorizable diagram contribution is suppressed due to the cancellation of Wilson coefficients $C_1 + C_2/3$. After the inclusion of the vertex corrections, the factorizable diagram
contributions in Fig.~\ref{fig:fig1}(a) and (b) become comparable with the nonfactorizable ones in Fig.~\ref{fig:fig1}(c) and (d).

In our numerical calculations for the branching ratios and polarization fractions,
the first two theoretical errors come from the Gegenbauer moments in the longitudinal and transverse twist-2 kaon-pion DAs,
namely, $a_{1K^*}^{||}=0.2\pm0.2, a_{2K^*}^{||}=0.5\pm0.5$ and $a_{1K^*}^{\perp}=0.3\pm0.3, a_{2K^*}^{\perp}=0.8\pm0.8$, respectively.
It is worthwhile to stress that for the hadronic charmonium $B$ decays, the energy release  may not be high enough for 
 justifying the PQCD leading order (LO) calculation,  and the theoretical accuracy  needs to be improved.
 Here, the significant vertex corrections are included, so that
the Gegenbauer momenta $a_{1K^*}^{\parallel}, a_{2K^*}^{\parallel}$ in Eq.~(\ref{eqphi0}) are redefined and different from those in our previous work~\cite{epjc79-37}, for which the hard kernels are evaluated only up to the LO level.
Therefore, a wide variation of the Gegenbauer moments are considered for the error estimation, such as $a_{1K^*}^{||}=0.2\pm0.2, a_{2K^*}^{||}=0.5\pm0.5$, which covers the previously determined central value $a_{1K^*}^{||}=0.05, a_{2K^*}^{||}=0.15$.
The third theoretical uncertainty results from the shape parameter $\omega_{B_{(s)}}$ of the $B_{(s)}$ meson distribution amplitude.
We adopt the value $\omega_B=0.40\pm0.04$~GeV or $\omega_{B_s}=0.50\pm0.05$~GeV and vary its value within a 10\% range,
which is supported by intensive PQCD studies~\cite{prd63-054008,epjc23-275,prd63-074009,plb504-6}.
The fourth error is caused by the variation of the hard scale $t$ from $0.75t$ to $1.25t$ (without changing $1/b_i$),
which characterizes the effect of the high order QCD contributions.
The last one is due to the hadronic parameter $\omega_c=0.60\pm0.06$ ($\omega_c=0.20\pm0.02$)~GeV for $J/\psi$ ($\psi(2S)$) meson ~\cite{epjc77-610,epjc60-107} from the wave functions of charmonium.

It is shown that the main uncertainties in our approach come from the Gegenbauer moments as listed in Table~\ref{tab:br} and Table~\ref{tab:br2}, which can reach about 60\% in magnitude totally.
The scale-dependent uncertainty is less than $25\%$ due to the inclusion of the NLO vertex corrections.
We have checked the sensitivity of our results to the choice of the shape parameter $\omega_c$ in the charmonia meson wave function.
The variation of $\omega_c$ will result in small changes of the branching ratio and polarization fractions, say less than $10\%$.
We have also examined the sensitivity of our results to the choices of other Gegenbauer moments ($a_{1s}^{\perp}, a_{1t}^{\perp},
a_{1a}^{\parallel}, a_{1v}^{\parallel}$) in the twist-3 DAs.
These Gegenbauer moments in the twist-3 DAs have a smaller impact on the total branching ratios than those in the twist-2 DAs.
With the increase (decrease) of $a_{1s}^{\perp}, a_{1t}^{\perp}, a_{1a}^{\parallel}, a_{1v}^{\parallel}$, the total branching ratios and the longitudinal polarization fractions become larger for the $B^0$ ($B^0_s$) decay modes.
The opposite pattern between the $B$ and $B_s$ modes can be understood because they decay into different $K\pi$  pairs.
The positive $a_1^{\perp,\parallel}$ related to a $K^+\pi^-$ pair carries an $\bar s$ quark,
while $a_1^{\perp,\parallel}$ should change the sign for $K^-\pi^+$ one with an $s$ quark.
The possible errors due to the uncertainties of $m_c$ and CKM matrix elements are very small and can be neglected safely.

It is observed that our predictions of the branching ratios for the involved $J/\psi$ channels agree well with the available data~\cite{pdg2018} in Table~\ref{tab:br}.
For the $B^0 \to \psi(2S) (K^{*0} \to)K^+\pi^-$ decay, the PQCD prediction for its branching ratio is well consistent
with the world average $(3.93\pm0.27) \times 10^{-4}$ within errors.
While for the $B^0_s \to \psi(2S) \bar{K}^{*0} \to \psi K^-\pi^+$ decay process, the central value of our theoretical prediction for its branching ratio is slightly smaller than that of the PDG number~\cite{pdg2018} within errors.
However, the PDG result is obtained by multiplying the best value $\mathcal {B}(B^{0}\to \psi(2S) K^+\pi^-)$ with
the measured ratio $\mathcal {B}(\bar{B}_s^{0}\to \psi(2S) K^+\pi^-)/\mathcal {B}(B^{0}\to\psi(2S) K^+\pi^-)$ via an intermediate state $K^*(892)^0$ from LHCb collaboration~\cite{plb747-484}.
We hope the experiment will provide a direct measurement to this decay mode in the future.

The two-body branching fraction ${\cal B}(B \to \psi K^{*0})$ can be extracted from the corresponding quasi-two-body decay modes in Table~\ref{tab:br} and~\ref{tab:br2} under the narrow width approximation relation
\begin{eqnarray}
\mathcal{B}(B \to \psi K^{*0}\to \psi K^+ \pi^- ) &=&
\mathcal{B}( B \to \psi K^{*0}) \cdot {\mathcal B}(K^{*0} \to K^+\pi^-),\label{eq:def1}
\end{eqnarray}
where we assume the $K^{*0}\to K\pi$ branching fraction to be 100\%.
The isospin conservation is assumed for the strong decays of an $I=1/2$ resonance $K^{*0}$ to $K\pi$ when we compute the branching fractions of the quasi-two-body process $B \to \psi K^{*0}\to \psi K^+\pi^-$, namely,
\begin{eqnarray}
\frac{\Gamma(K^{*0} \to K^+\pi^-)}{\Gamma(K^{*0} \to K\pi)}=2/3, ~\quad
\frac{\Gamma(K^{*0} \to K^0\pi^0)}{\Gamma(K^{*0} \to K\pi)}=1/3. \label{eq:def2}
\end{eqnarray}
When compared with previous theoretical predictions for $B \to \psi K^*$ in the two-body framework both in the PQCD approach
~\cite{prd71-114008,prd89-094010,epjc77-610} and in the QCDF approach ~\cite{prd65-094023},
one can find that the branching ratios of the quasi-two-body decay modes are in good agreement with those two-body analyses
based on the PQCD approach ~\cite{prd89-094010,epjc77-610}.
Taking $B^0 \to J/\psi K^{*0}$ decay as an example, we obtain the ${\cal B}( B^0 \to J/\psi K^{*0}) \approx 1.25 \times 10^{-3}$ from the value
as listed in  the first section of Table~\ref{tab:br},  which agrees well with the theoretical prediction
${\cal B}( B^0 \to J/\psi K^{*0})=(1.23^{+0.42}_{-0.36}) \times 10^{-3}$ as given in Ref.~\cite{prd89-094010}, and with the world average
of  the measured ones from {\it BABAR}, CDF and Belle Collaborations \cite{prd58-072001, prd90-112009,prl94-141801}:
$(1.3\pm 0.06)  \times 10^{-3}$ from HFLAV~\cite{1909-12524}.
The consistency of the PQCD predictions for the branching ratios supports the usability of the PQCD factorization for exclusive hadronic $B$ meson decays.
Our PQCD predictions for the branching ratios are also consistent  with those in the QCDF approach~\cite{prd65-094023} within errors.

\begin{table}[t]
\caption{ The PQCD predictions  for the branching ratios and polarization fractions defined in Eq.~(\ref{pol}) of the $P$-wave resonance channels in
the  $B^0_{(s)} \to J/\psi K^{\pm} \pi^{\mp}$ decay together with experimental data~\cite{pdg2018}.
The theoretical errors are attributed to the variation of the longitudinal Gegenbauer
moments ($a_{1K^*}^{||}$ and $a_{2K^*}^{||}$) and transverse ones ($a_{1K^*}^{\perp}$ and $a_{2K^*}^{\perp}$),
the shape parameters $\omega_{B_{(s)}}$ in the wave function of $B_{(s)}$ meson and the hard scale $t$, and the parameters in the wave functions of charmonium, respectively.}
\label{tab:br}
\begin{ruledtabular}
\begin{tabular}[t]{cccc}
Modes &\qquad & PQCD predictions  & Experiment \footnotemark[1] \\
\hline
$B^0 \to J/\psi (K^{*}(892)^{0} \to)K^+\pi^- $ &$\mathcal{B}(10^{-4})$ &$8.31^{+3.34+2.32+1.89+1.81+0.00}_{-2.41-1.77-1.78-1.22-0.14}$ &$8.47\pm0.03$ \\
& $f_0(\%)$  &$55.9^{+13.4+14.2+0.0+1.8+0.0}_{-16.6-12.9-0.5-0.9-0.1}$ &$57.1\pm0.7$\\
 & $f_{\parallel}(\%)$ &$21.1^{+8.0+7.2+0.1+0.3+0.1}_{-6.4-8.2-0.1-1.0-0.1}$&$-$ \\
&$f_{\perp}(\%)$ &$23.0^{+8.7+5.6+0.4+0.6+0.1}_{-6.9-6.2-0.0-0.8-0.0}$&$21.1\pm0.8$ \\
\hline
$B^0 \to J/\psi (K^{*}(1410)^{0} \to)K^+\pi^- $&$\mathcal{B}(10^{-5})$ &$1.98^{+0.72+0.52+0.33+0.68+0.02}_{-0.46-0.24-0.28-0.38-0.02}$ &$-$\\
& $f_0(\%)$ &$53.1^{+13.2+11.3+2.3+2.9+0.6}_{-13.3-11.6-0.8-2.1-0.2}$ &$-$\\
& $f_{\parallel}(\%)$ &$20.1^{+5.8+8.3+0.7+0.4+0.2}_{-5.8-8.7-1.1-1.1-0.7}$&$-$ \\
&$f_{\perp}(\%)$ &$26.8^{+7.6+3.5+0.1+1.7+0.1}_{-7.6-2.8-1.1-1.2-0.1}$ &$-$\\
\hline
$B^0 \to J/\psi (K^{*}(1680)^{0} \to)K^+\pi^- $ &$\mathcal{B}(10^{-5})$ &$2.02^{+0.66+0.54+0.36+0.71+0.02}_{-0.67-0.54-0.52-0.62-0.03}$&$-$ \\
& $f_0(\%)$ &$51.2^{+12.9+10.8+1.4+2.9+1.4}_{-14.4-11.4-0.9-4.0-0.2}$ &$-$ \\
& $f_{\parallel}(\%)$ &$19.7^{+5.8+10.5+0.7+1.8+0.5}_{-5.2-9.4-0.2-0.8-0.3}$ &$-$\\
&$f_{\perp}(\%)$ &$29.1^{+8.7+1.5+0.2+2.2+0.0}_{-7.7-2.4-1.2-2.1-1.1}$&$-$ \\
\hline
$B^0_s \to J/\psi (\bar{K}^{*}(892)^{0} \to)K^-\pi^+ $&$\mathcal{B}(10^{-5})$ &$2.29^{+1.26+0.87+0.79+0.50+0.05}_{-0.74-0.58-0.52-0.27-0.00}$&$2.73\pm0.27$\\
& $f_0(\%)$ &$54.3^{+17.5+17.6+1.0+1.9+1.7}_{-19.7-15.9-0.0-0.0-0.0}$&$49.7\pm3.5$\\
& $f_{\parallel}(\%)$ &$24.7^{+10.6+8.8+0.1+0.0+0.0}_{-9.5-9.5-0.1-0.9-0.6}$ &$17.9\pm3.0$ \\
&$f_{\perp}(\%)$ &$21.0^{+8.9+7.1+0.0+0.0+0.0}_{-8.1-7.3-0.9-1.0-1.1}$ &$-$\\
\hline
$B^0_s \to J/\psi (\bar{K}^{*}(1410)^{0} \to)K^-\pi^+ $ &$\mathcal{B}(10^{-7})$
&$7.22^{+2.64+2.24+1.56+1.78+0.00}_{-2.71-1.28-1.43-1.20-0.02}$&$-$\\
& $f_0(\%)$ &$48.3^{+14.7+10.4+0.3+1.4+0.3}_{-15.9-12.1-0.0-1.4-0.0}$ &$-$\\
& $f_{\parallel}(\%)$&$27.8^{+8.6+8.7+0.0+0.0+0.2}_{-7.9-8.7-0.2-0.6-0.3}$ &$-$\\
&$f_{\perp}(\%)$&$23.9^{+7.4+3.6+0.2+1.5+0.0}_{-6.7-2.1-0.1-0.7-0.2}$ &$-$\\
\hline
$B^0_s \to J/\psi (\bar{K}^{*}(1680)^{0} \to)K^-\pi^+ $ &$\mathcal{B}(10^{-7})$
&$7.36^{+2.44+2.37+1.29+2.18+0.03}_{-1.46-1.27-1.38-1.05-0.04}$ &$-$\\
& $f_0(\%)$ &$42.6^{+14.8+7.6+0.0+2.4+0.2}_{-13.5-11.1-0.9-0.8-0.4}$ &$-$\\
& $f_{\parallel}(\%)$ &$29.6^{+7.2+10.5+0.9+0.0+0.5}_{-7.7-8.0-0.1-0.6-0.6}$ &$-$\\
&$f_{\perp}(\%)$ &$27.8^{+6.6+2.2+0.1+0.4+0.1}_{-7.2-1.0-0.1-1.3-0.1}$&$-$\\
\end{tabular}
\end{ruledtabular}
\footnotetext[1]{ The experimental results are obtained by multiplying the relevant measured two-body branching
ratios according to the Eq.~(\ref{eq:def1}). }
\end{table}

\begin{figure}[tbp]
\centerline{\epsfxsize=8cm \epsffile{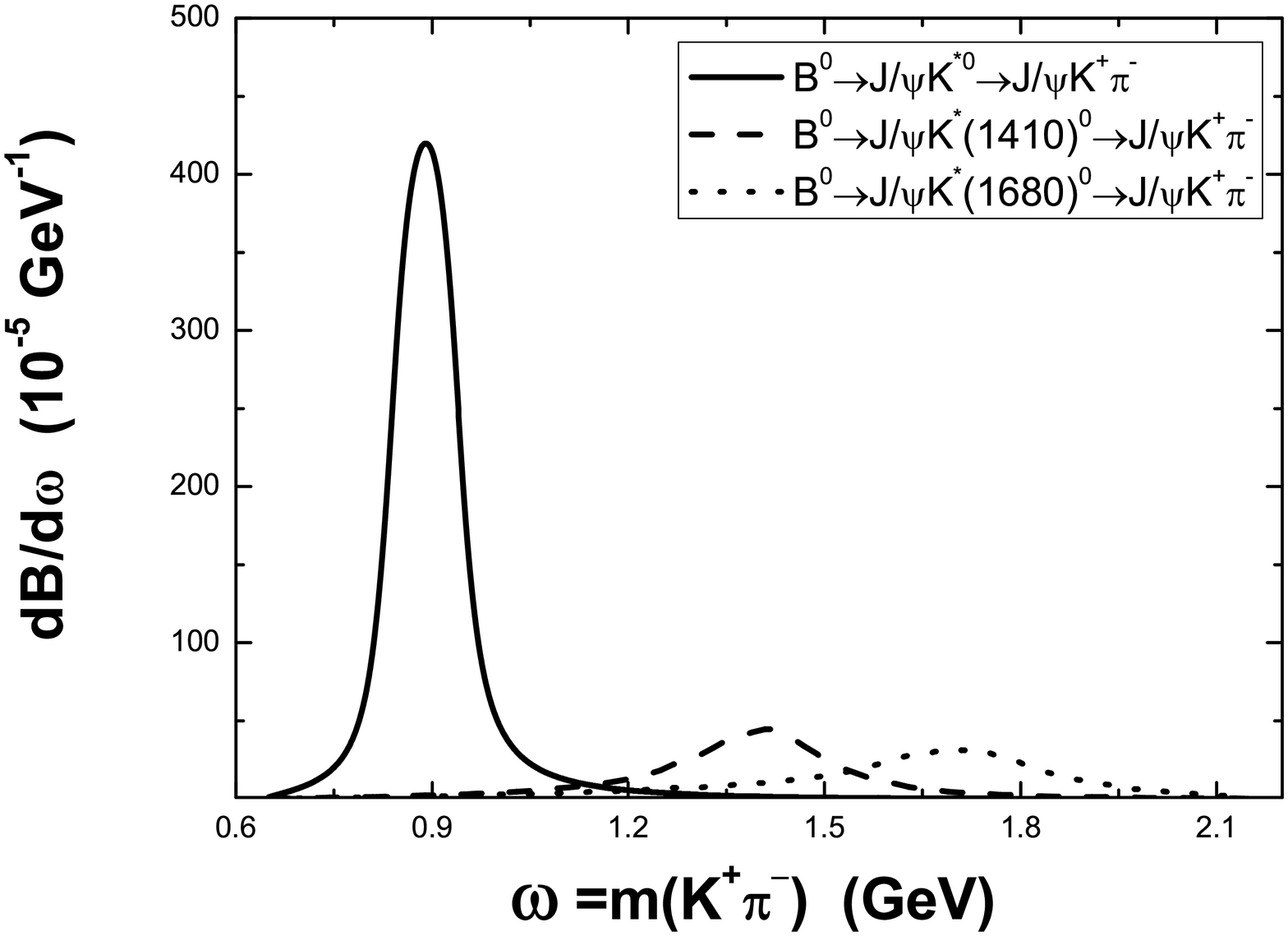}
            \epsfxsize=8cm \epsffile{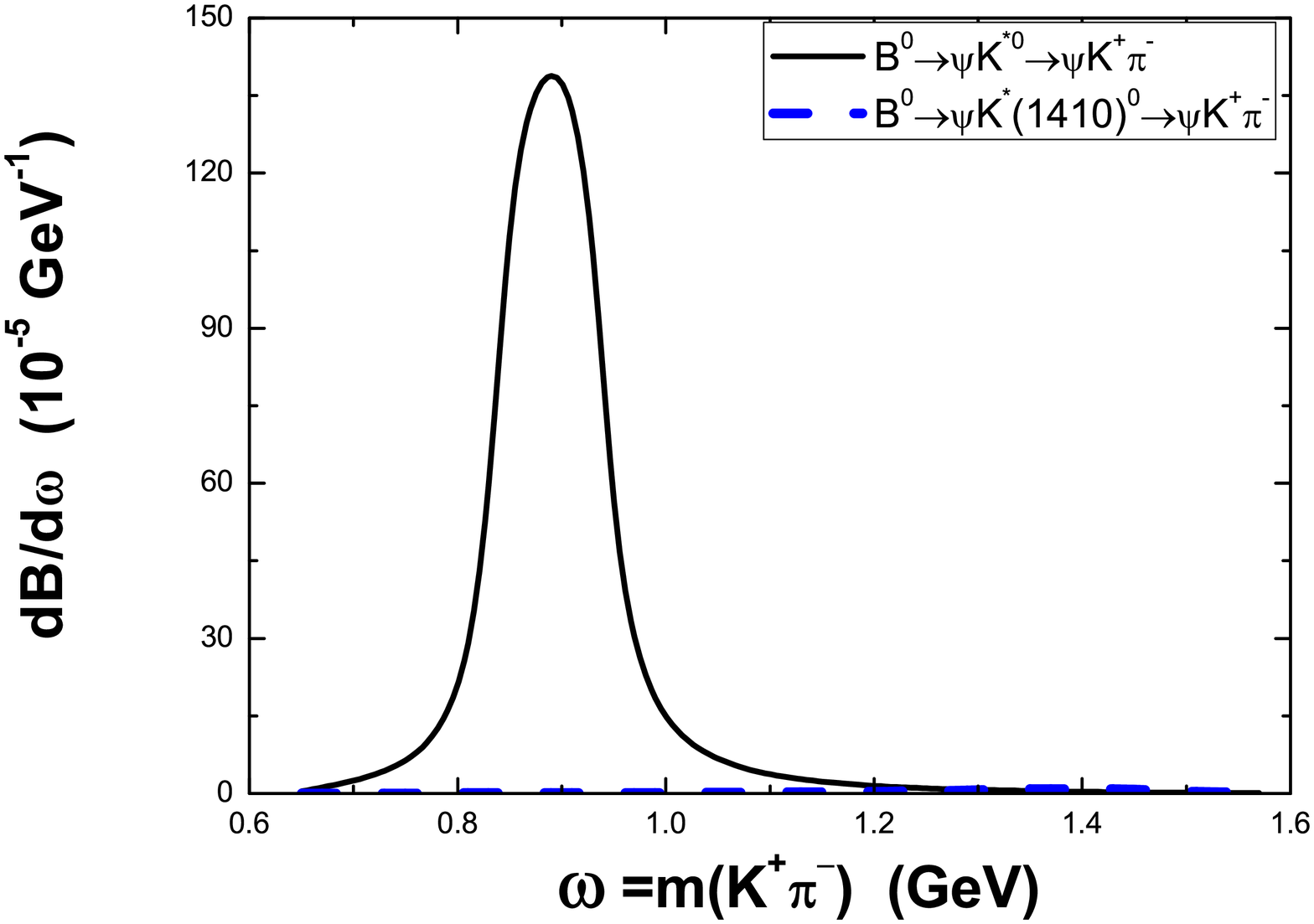}}
\vspace{-0.2cm}
  {\scriptsize\bf (a)\hspace{7.5cm}(b)}
\caption{(a) Differential branching ratios for the $B^0\to J/\psi [K^{*0},K^*(1410)^0, K^*(1680)^0\to]K^+\pi^-$ decays,
         and (b)Differential branching ratios for the $B^0\to \psi(2S) [K^{*0},K^*(1410)^0 \to]K^+\pi^-$ decays.}
\label{fig-br}
\end{figure}

\begin{table}
\caption{  The PQCD predictions for the branching ratios and polarization fractions defined in Eq.~(\ref{pol}) of the $P$-wave resonance channels in the
$B^0_{(s)} \to \psi(2S) K^{\pm} \pi^{\mp}$ decay together with experimental data~\cite{pdg2018}.
The theoretical errors are attributed to the variation of the longitudinal Gegenbauer
moments ($a_{1K^*}^{||}$ and $a_{2K^*}^{||}$) and transverse ones ($a_{1K^*}^{\perp}$ and $a_{2K^*}^{\perp}$),
the shape parameters $\omega_{B_{(s)}}$ in the wave function of $B_{(s)}$ meson and the hard scale $t$, and the hard scale $t$, and the parameters in the wave functions of charmonium, respectively.}
\label{tab:br2}
\begin{ruledtabular}
\begin{tabular}[t]{cccc}
Modes &\qquad & PQCD predictions  & Experiment \footnotemark[1] \\
\hline
$B^0 \to \psi(2S)(K^{*}(892)^{0} \to)K^+\pi^- $ &$\mathcal{B}(10^{-4})$
&$3.38^{+1.03+0.97+0.89+0.79+0.03}_{-0.84-0.74-0.77-0.58-0.01}$&$3.93\pm0.27$\\
& $f_0(\%)$ &$46.4^{+18.7+12.4+0.0+1.1+0.2}_{-16.6-10.4-1.0-2.6-0.0}$&$46.3^{+2.8}_{-4.0}$\\
& $f_{\parallel}(\%)$ &$25.4^{+7.9+7.5+0.5+1.4+0.0}_{-6.3-8.5-0.2-0.7-0.3}$&$-$ \\
&$f_{\perp}(\%)$ &$28.2^{+8.8+3.5+0.5+1.2+0.0}_{-6.9-4.0-0.0-0.4-0.1}$&$30.0\pm6.0$\\
\hline
$B^0 \to \psi(2S)(K^{*}(1410)^{0} \to)K^+\pi^- $ &$\mathcal{B}(10^{-6})$
&$4.88^{+1.41+1.33+1.08+1.85+0.09}_{-1.11-0.93-0.81-1.02-0.01}$ &$-$\\
&$f_0(\%)$ &$44.8^{+13.3+10.1+1.1+3.0+0.9}_{-15.2-10.2-1.4-2.9-0.5}$ &$-$\\
&$f_{\parallel}(\%)$ &$23.2^{+7.3+10.2+1.2+1.1+0.6}_{-5.5-10.6-0.8-1.3-0.5}$ &$-$\\
&$f_{\perp}(\%)$ &$32.0^{+8.8+1.3+0.3+1.8+0.0}_{-7.7-0.7-0.3-1.7-0.3}$&$-$\\
\hline
$B^0_s \to \psi(2S) (\bar{K}^{*}(892)^{0} \to)K^-\pi^+ $ &$\mathcal{B}(10^{-6})$
&$7.69^{+3.02+3.32+1.53+1.80+0.05}_{-1.93-2.26-1.75-1.21-0.00}$&$22.0\pm3.3$ \\
&$f_0(\%)$ &$39.4^{+17.7+14.9+0.8+1.6+0.2}_{-18.9-12.7-1.3-2.0-0.5}$&$52.0\pm6.0$\\
& $f_{\parallel}(\%)$&$32.0^{+10.0+8.2+0.8+0.7+0.3}_{-9.4-10.4-0.5-0.9-0.8}$ &$-$\\
&$f_{\perp}(\%)$ &$28.6^{+8.9+4.5+0.5+1.3+1.2}_{-8.3-4.6-0.3-0.7-0.5}$ &$-$\\
\hline
$B^0_s \to \psi(2S) (\bar{K}^{*}(1410)^{0} \to)K^-\pi^+ $ &$\mathcal{B}(10^{-7})$
&$1.50^{+0.51+0.54+0.37+0.47+0.00}_{-0.30-0.36-0.29-0.20-0.01}$ &$-$\\
&$f_0(\%)$ &$28.5^{+17.8+8.1+0.5+2.5+0.8}_{-17.3-8.2-0.0-1.3-0.0}$ &$-$\\
&$f_{\parallel}(\%)$&$34.1^{+8.3+10.7+0.2+0.1+0.1}_{-8.5-11.2-0.1-0.8-1.0}$ &$-$\\
&$f_{\perp}(\%)$ &$37.4^{+8.9+3.2+0.0+2.1+0.7}_{-9.3-2.6-0.6-2.7-0.9}$&$-$\\
\end{tabular}
\end{ruledtabular}
\footnotetext[1]{ The experimental results are obtained by multiplying the relevant measured two-body branching ratios according to the Eq.~(\ref{eq:def1}). }
\end{table}
In Fig.~\ref{fig-br}(a), we show the $\omega$-dependence of the differential decay rate $d{\cal B}(B^0 \to J/\psi K^+\pi^-)/d\omega$
after the inclusion of the possible contributions from the resonant states $K^*$ (the solid curve), $K^*(1410)$ (the dashed curve), $K^*(1680)$ (the dotted curve).
Similarly, we display the PQCD prediction for $d{\cal B}/d\omega$ for $B^0 \to \psi(2S)K^{*0}\to \psi(2S)K^+\pi^-$ (the solid curve) and $B^0 \to \psi(2S)K^*(1410)^0\to \psi(2S)K^+\pi^-$ (the blue dashed curve) in Fig.~\ref{fig-br}(b).
For the considered decay modes $B^0 \to \psi K^+\pi^-$, the dynamical limit on the value of invariant mass $\omega$ is $(m_{K^+}+m_{\pi^-}) \leq \omega \leq (m_B-m_{\psi})$.
For $B^0 \to \psi(2S) K^+\pi^-$ decays, since $m_{K^*(1680)} > \omega_{max}=(m_B-m_{\psi(2S)})$, the resonance $K^*(1680)$ can not contribute to this decay.
Obviously, the differential branching ratios of these decays exhibit peaks at the pole mass of the resonant states.
Thus, the main portion of the branching ratios lies in the region around the resonance as expected.
For $B^0 \to J/\psi K^{*0} \to J/\psi K^+\pi^-$ decay, the central values of the branching ratio ${\cal B}$ are $4.29\times10^{-4} $ and $6.49\times 10^{-4}$
when the integration over $\omega$ is limited in the range of
$\omega=[m_{K^*}-0.5\Gamma_{K^*}, m_{K^*}+0.5\Gamma_{K^*}]$ or
$\omega=[m_{K^*}-\Gamma_{K^*}, m_{K^*}+\Gamma_{K^*}]$ respectively, which amount to
$51.6\%$ and $78.1\%$ of the total branching ratio ${\cal B}=8.31\times10^{-4}$ as listed in Table~\ref{tab:br}.
The peak of $K^*(1680)$ has slightly smaller strength than the $K^*(1410)$, while its broader width compensates the integrated strength over the whole phase space.
Therefore, the branching ratios of the two components are of a comparable size as predicted in our work.

From the numerical results as given in Table~\ref{tab:br} and \ref{tab:br2}, we predict the relative ratio $R_1$ between the branching ratios of $B$
meson decays involving $\psi(2S)$ and $J/\psi$ with the resonance $K^*(1410)^{0}$,
\begin{eqnarray}
R_1(K^*(1410))&=&\frac{{\cal B}(B^0 \to \psi(2S)(K^{*}(1410) ^{0}\to) K^+\pi^-)}{{\cal B}(B^0 \to J/\psi(K^{*}(1410) ^{0} \to) K^+\pi^-)}=0.25^{+0.01}_{-0.03},
\end{eqnarray}
which is smaller than the corresponding ratio of $K^*$ reported by the LHCb measurement~\cite{epjc72-2118},
\begin{eqnarray}
R_1^{exp} = \frac{{\cal B}(\bar{B}^0 \to \psi(2S) K^{*0})}{{\cal B}(\bar{B}^0 \to J/\psi K^{*0})}
= 0.476 \pm 0.014 \pm0.010 \pm 0.012.
\end{eqnarray}
The gap is governed by the different masses and widths in the Breit-Wigner functions of the $K^*$ and $K^*(1410)$.
It is expected that the forthcoming LHCb and Belle-II experiment to provide a direct measurement of $R_1(K^*(1410))$.

The polarization fractions defined in Eq.~(\ref{pol}) associated with the available data are also listed  in Tables~\ref{tab:br} and~\ref{tab:br2},
which have the same origin of theoretical uncertainties as the branching ratios.
For these decays, the contributions from the non-factorizable tree diagrams in Fig.~\ref{fig:fig1}(c,d) are comparable with those of the color-suppressed tree
diagrams though the latter are enhanced by the involving vertex corrections.
It is easy to see that the fraction of the longitudinal polarization can be generally reduced to about $\sim 50\%$,
while the parallel and perpendicular ones are roughly equal.
Especially for $B_s\rightarrow \psi(2S)$ modes, all the three transversity amplitudes are of comparable size.
The results are quite different from the expectation in the factorization assumption that the longitudinal
polarization should dominate based on the quark helicity analysis~\cite{zpc1-269,prd64-117503}.
We can also adopt the helicity amplitudes $(\mathcal {A}_{0},\mathcal {A}_{+},\mathcal {A}_{-})$, which
are related to the spin amplitudes $(\mathcal {A}_{0},\mathcal {A}_{\parallel},\mathcal {A}_{\perp})$ in Appendix by
\begin{eqnarray}
\label{eq:helicity}
\mathcal {A}_{\pm}=\frac{\mathcal {A}_{\parallel}\pm\mathcal {A}_{\perp}}{\sqrt{2}},
\end{eqnarray}
while $\mathcal {A}_{0}$ is common to both bases.
From the results of Tables~\ref{tab:br} and~\ref{tab:br2} and Eq.~(\ref{eq:helicity}),
one have a hierarchy of helicity amplitudes $|\mathcal {A}_0|\sim|\mathcal {A}_+|>|\mathcal {A}_{-}|$,
which are very similar to the cases of two-body charmonium $B$ decays \cite{epjc77-610}.

The PQCD predictions for the polarization fractions of most considered $
B^0_{(s)} \to \psi K^* \to \psi K\pi$ agree with currently available data within errors.
For $B^0_s \to \psi(2S) (\bar{K}^{*}(892)^{0} \to)K^-\pi^+ $ decay, although the central value of
$f_0\approx 39.4\%$ is a little smaller than the measured one  $f_0^{exp}=52\%$, but they are consistent with each other
due to the still large theoretical and experimental uncertainties.
As stressed above, we expect  a systematic angular analysis of  the $B^0_s \to \psi(2S) (\bar{K}^{*0} \to)K^-\pi^+$ decay mode
to accurately extract various polarization amplitudes.
According to Table~II and Eq.~(9) from Belle collaboration~\cite{prd90-112009},
the central values of longitudinal, parallel polarization fractions are $46.3\%$ and $24.8\%$ for $\bar{B}^0 \to J/\psi K^*(1410)^0 \to J/\psi K^-\pi^+$,
as well as $37.2\%$ and $31.9\%$ for $\bar{B}^0 \to J/\psi K^*(1680)^0 \to J/\psi K^-\pi^+$.
In comparison to our predictions in Tables~\ref{tab:br} and~\ref{tab:br2},
their longitudinal polarizations are small but the parallel ones are large.
As mentioned before, since the tensor and vector decay constants for $K^*(1410)$ or $K^*(1680)$ are still not known yet,
we use the decay constants of $K^*$ to define the ratio $F_{K\pi}^{\perp}(s)/F_{K\pi}^{\parallel}(s)$.
In fact, the ratios $F_{K\pi}^{\perp}(s)/F_{K\pi}^{\parallel}(s)$ should be regarded as free parameters and determined by fitting to the data in the absence of any theoretical and experimental bases.
However, the statistical error in~\cite{prd90-112009} are large and the corresponding systematic error are still absent,
it is not possible to perform a global fitting from the current data.
More precise measurements of such decay channels are expected to help us to test and improve our theoretical calculations.

\section{CONCLUSION}

In this work, we have studied the quasi-two-body decays $B^0_{(s)} \to \psi [K^{*0}, K^*(1410)^0, K^*(1680)^0$ $\to] K\pi$ in the PQCD
factorization approach by introducing the kaon-pion DAs.
Analogous  to the case of the $P$-wave pion pair in the final state, the kaon-pion DAs corresponding to the longitudinal and transverse polarizations are constructed through a perturbative evaluation of the hadronic matrix elements associated with the various spin projectors.

The SU(3) flavor symmetry breaking term shows an important effect on the longitudinal polarizations of the kaon-pion pair.
Besides, we also fixed the hadronic parameters involved in the kaon-pion DAs on basis of the data for the
branching ratios and polarization fractions of the relevant decay modes.

Our PQCD predictions for  the branching ratios and polarization fractions for the considered $B^0 \to \psi (K^* \to) K\pi$ decays are in good agreement with the existing data.
The clear differences  between the PQCD predictions and the measured values for the $B^0_s \to \psi(2S) (\bar{K}^{*0} \to) K^-\pi^+$ decay could be tested by the forthcoming experiments.
The branching ratios of the two-body $B \to \psi K^*$ can be extracted from the corresponding quasi-two-body modes
by employing the narrow width approximation.

We also calculated the branching ratios and polarization fractions of the $B^0_{(s)} \to \psi[K^*(1410),$ $K^*(1680)\to] K\pi$ decays
, and defined the new ratio $R_1(K^*(1410))$ among the branching ratios of the considered decay modes,
these predictions could be tested by the future experimental measurements.

\begin{acknowledgments}
Many thanks to Hsiang-nan Li and Wen-Fei Wang for valuable discussions.
This work was supported by the National Natural Science Foundation of China under the No.~11947013, No.~11605060, No.~11775117, and No.~11547020.
Ya Li is also supported by the Natural Science Foundation of Jiangsu Province under Grant No.~BK20190508 and the Research Start-up Funding of Nanjing
Agricultural University.
Zhou Rui is supported in part by the Natural Science Foundation of Hebei Province under Grant No.~A2019209449.

\end{acknowledgments}

\appendix
\section{Decay amplitudes}
The contributions from the longitudinal polarization, the normal polarization, and the transverse polarization are marked by the subscripts $L$, $N$ and $T$, respectively.
The superscript $LL$, $LR$, and $SP$  refers to the contributions from $(V-A)\otimes(V-A)$, $(V-A)\otimes(V+A)$ and $(S-P)\otimes(S+P)$ operators, respectively.
The total decay amplitude is decomposed into
\begin{eqnarray}\label{eq:alnt}
\mathcal{A}=\mathcal{A}_L+\mathcal{A}_N \epsilon_{T}\cdot \epsilon_{3T}
+i \mathcal{A}_T \epsilon_{\alpha\beta\rho\sigma} n_+^{\alpha} n_-^{\beta} \epsilon_{T}^{\rho} \epsilon_{3T}^{\sigma},
\end{eqnarray}
where the three individual polarization amplitudes are written as
\begin{eqnarray}
\mathcal{A}_{L,N,T}(B^0_{(s)}\to \psi K\pi)&=&\frac{G_F}{\sqrt{2}}\Big\{V^*_{cb}V_{cs(cd)}
\Big [(C_1+\frac{1}{3}C_2)\mathcal{F}_{L,N,T}^{LL}+C_2\mathcal{M}_{L,N,T}^{LL} \Big]\nonumber\\
&&-V^*_{tb}V_{ts(td)}\Big [(C_3+\frac{1}{3}C_4+C_9+\frac{1}{3}C_{10})\mathcal{F}_{L,N,T}^{LL}\non
&& +(C_5+\frac{1}{3}C_6+C_7+\frac{1}{3}C_{8})\mathcal{F}_{L,N,T}^{LR}\nonumber\\
&& + (C_4+C_{10})\mathcal{M}_{L,N,T}^{LL}+(C_6+C_8)\mathcal{M}_{L,N,T}^{SP}\Big ]\Big\},
\end{eqnarray}
with the CKM matrix elements $V_{ij}$ and the Fermi coupling constant $G_F$.
The Wilson coefficients $C_i$ encode the hard dynamics of weak decays.
The above amplitudes are related to those in Eq.~(\ref{eq:dfenzhibi}) via
\begin{eqnarray}
\mathcal{A}_0=\mathcal{A}_L, \quad \mathcal{A}_{\parallel}=\sqrt{2}\mathcal{A}_{N},
\quad \mathcal{A}_{\perp}=\sqrt{2}\mathcal{A}_{T}.
\end{eqnarray}
The explicit amplitudes $\mathcal{F(M)}$ from the factorizable (nonfactorizable) diagrams in Fig.~\ref{fig:fig1} can be obtained straightforwardly just by replacing the twist-2 or twist-3 DAs of the $\pi\pi$ system with the corresponding twists of the $K\pi$ ones in Eqs.~({\ref{eqphi0}})-(\ref{eqphiv}), since the $P$-wave kaon-pion distribution amplitude in Eq.~(\ref{eq:phifunc}) has the same Lorentz structure as that of two-pion ones in Ref.~\cite{prd98-113003}.



\begin{thebibliography}{199}
\bibitem{prd71-032005}
B.~Aubert et al., [BABAR Collaboration],  Phys. Rev. D {\bf 71}, 032005 (2005).

\bibitem{prl94-141801}
B.~Aubert et al., [BABAR Collaboration], Phys. Rev. Lett. {\bf 94}, 141801 (2005).

\bibitem{prd76-031102}
B.~Aubert et al., [BABAR Collaboration], Phys. Rev. D {\bf 76}, 031102 (2007).

\bibitem{plb538-11}
K. Abe et al. [Belle Collaboration], \plb {\bf 538}, 11 (2002).

\bibitem{prl95-091601}
R.~Itoh et al., [Belle Collaboration], Phys. Rev. Lett. {\bf 95}, 091601 (2005).

\bibitem{prd80-031104}
R. Mizuk et al., [Belle Collaboration], Phys. Rev. D {\bf 80}, 031104(R) (2009).

\bibitem{prd88-072004}
R.~Itoh et al., [Belle Collaboration], Phys. Rev. D {\bf 88}, 072004 (2013).

\bibitem{prd90-112009}
K.~Chilikin et al., [Belle Collaboration], Phys. Rev. D {\bf 90}, 112009 (2014).
\bibitem{epjc72-2118}
R.~Aaij et al., [LHCb Collaboration], \epjc {\bf 72}, 2118 (2012).
\bibitem{prd88-052002}
R.~Aaij et al., [LHCb Collaboration], Phys. Rev. D {\bf 88}, 052002 (2013).
\bibitem{jhep11-082}
R.~Aaij et al., [LHCb Collaboration], \jhep {\bf 11}, 082 (2015).
\bibitem{plb747-484}
R.~Aaij et al., [LHCb Collaboration], \plb {\bf 747}, 484 (2015).
\bibitem{prl76-2015}
F.~Abe et al., [CDF Collaboration], Phys. Rev. Lett. {\bf 76}, 2015 (1996).
\bibitem{prd58-072001}
F.~Abe et al., [CDF Collaboration], Phys. Rev. D {\bf 58}, 072001 (1998).
\bibitem{prl85-4668}
T.~Affolder et al., [CDF Collaboration], Phys. Rev. Lett. {\bf 85}, 4668(2000).
\bibitem{prl94-101803}
D.~Acosta et al., [CDF Collaboration], Phys. Rev. Lett. {\bf 94}, 101803 (2005).
\bibitem{prd83-052012}
T.~Aaltonen et al., [CDF Collaboration], Phys. Rev. D {\bf 83}, 052012 (2011).
\bibitem{prl79-4533}
C.P.~Jessop et al., [CLEO Collaboration], Phys. Rev. Lett. {\bf 79}, 4533 (1997).

\bibitem{prl102-032001}
V.M.~Abazov et al., [D0 Collaboration], Phys. Rev. Lett. {\bf 102}, 032001 (2009).

\bibitem{prd89-094013}
I.~Bediaga, T.~Frederico, O.~Louren\c{c}o, \prd  {\bf 89}, 094013 (2014).
\bibitem{1512-09284}
I.~Bediaga, P.C.~Magalh\~{a}es, arXiv:1512.09284 [hep-ph].
\bibitem{prd89-053015}
X.W.~Kang, B.~Kubis, C.~Hanhart, U.G.~Mei\ss ner, \prd {\bf 89}, 053015 (2014).

\bibitem{prl83-1914}
M.~Beneke, G.~Buchalla, M.~Neubert, C.T.~Sachrajda, Phys. Rev. Lett.  {\bf 83}, 1914 (1999).
\
\bibitem{npb591-313}
M.~Beneke, G.~Buchalla, M.~Neubert, C.T.~Sachrajda, Nucl. Phys. B  {\bf 591}, 313 (2000).

\bibitem{npb606-245}
M.~Beneke, G.~Buchalla, M.~Neubert, C.T.~Sachrajda, Nucl. Phys. B   {\bf 606}, 245 (2001).

\bibitem{npb675-333}
M.~Beneke, M.~Neubert, Nucl. Phys. B {\bf 675}, 333 (2003).

\bibitem{npb899-247}
S.~Kr\"{a}nkl, T.~Mannel, J.~Virto, Nucl. Phys. B {\bf 899}, 247 (2015).

\bibitem{plb622-207}
A.~Furman, R.~Kami\'nski, L.~Le\'sniak, B.~Loiseau,  \plb  {\bf 622}, 207 (2005).

\bibitem{prd74-114009}
B.~El-Bennich, A.~Furman, R.~Kami\'nski, L.~Le\'sniak, B.~Loiseau, \prd {\bf 74}, 114009 (2006).

\bibitem{prd79-094005}
B.~El-Bennich, A.~Furman, R.~Kami\'nski, L.~Le\'sniak, B.~Loiseau, B.~Moussallam,  \prd  {\bf 79}, 094005 (2009).

\bibitem{APPB42-2013}
J.P.~Dedonder, A.~Furman, R.~Kami\'nski, L.~Le\'sniak, B.~Loiseau, Acta. Phys. Polon. B {\bf 42}, 2013 (2011).
\bibitem{prd76-094006}
H.Y.~Cheng, C.K.~Chua, A.~Soni,  \prd  {\bf 76}, 094006 (2007). 
\bibitem{prd88-114014}
H.Y.~Cheng, C.K.~Chua, \prd  {\bf 88}, 114014 (2013).
\bibitem{prd94-094015}
H.Y.~Cheng, C.K.~Chua, Z.Q.~Zhang,  \prd  {\bf 94}, 094015 (2016).
\bibitem{prd89-094007}
Y.~Li, \prd {\bf 89}, 094007 (2014).
\bibitem{prd87-076007}
Z.H.~Zhang, X.H.~Guo, Y.D.~Yang, \prd {\bf 87}, 076007 (2013).
\bibitem{jhep10-117}
R.~Klein, T.~Mannel, J.~Virtob, K.~Keri Vos, J. High Energy Phys. {\bf 10}, 117 (2017).
\bibitem{prd72-094031}
M.~Gronau, J.L.~Rosner,  \prd  {\bf 72}, 094031 (2005).
\bibitem{plb727-136}
M.~Gronau, \plb  {\bf 727}, 136 (2013).
\bibitem{prd72-075013}
G.~Engelhard, Y.~Nir, G.~Raz, \prd  {\bf 72}, 075013 (2005).
\bibitem{prd84-056002}
M.~Imbeault, D.~London,  \prd  {\bf 84}, 056002 (2011).
\bibitem{plb728-579}
D.~Xu, G.N.~Li, X.G.~He, \plb  {\bf 728}, 579 (2014).
\bibitem{prd91-014029}
X.G.~He, G.N.~Li, D.~Xu,  \prd  {\bf 91}, 014029 (2015).
\bibitem{prd63-074009}
C.D.~L\"u, K.~Ukai, M.Z.~Yang, \prd {\bf 63}, 074009 (2001).
\bibitem{plb504-6}
Y.Y.~Keum, H.N.~Li, A.I.~Sanda, \plb {\bf 504}, 6-14 (2001).
\bibitem{ppnp51-85}
H.N.~Li,  \ppnp {\bf 51}, 85 (2003) and references therein.
\bibitem{prd70-054015}
C.W.~Bauer, D.~Pirjol, I.Z.~Rothstein, I.W.~Stewart, \prd {\bf 70}, 054015 (2004).
\bibitem{prd74-034010}
C.W.~Bauer, I.Z.~Rothstein, I.W.~Stewart, \prd {\bf 74}, 034010 (2006).
\bibitem{npb692-232}
M.~Beneke, Y.~Kiyo, D.S.~Yang, \npb {\bf 692}, 232 (2004).
\bibitem{prd72-098501}
M.~Beneke, G.~Buchalla, M.~Neubert, C.T.~Sachrajda, \prd {\bf 72}, 098501 (2005).
\bibitem{prd72-098502}
C.W.~Bauer, D.~Pirjol, I.Z.~Rothstein, I.W.~Stewart, \prd {\bf 72}, 098502 (2005).
\bibitem{prd59-092004}
H.Y.~Cheng, K.C.~Yang, \prd {\bf 59}, 092004 (1999).
\bibitem{prd71-114008}
C.H.~Chen, H.N.~Li, \prd {\bf 71}, 114008 (2005).
\bibitem{prl74-4388}
H.N.~Li, H.L.~Yu, Phys. Rev. Lett. {\bf 74}, 4388 (1995).
\bibitem{plb348-597}
H.N.~Li, Phys. Lett. B {\bf 348}, 597 (1995).
\bibitem{JHEP06-013}
H.N.~Li, Y.M.~Wang,  \jhep {\bf 06}, 013 (2015).
\bibitem{JHEP02-008}
H.N.~Li, Y.L.~Shen, Y.M.~Wang, \jhep {\bf 02}, 008 (2013).

\bibitem{prd89-094010}
X.~Liu, W.~Wang, Y.H.~Xie, Phys. Rev. D {\bf 89}, 094010 (2014).
\bibitem{prd90-114030}
Z.~Rui, Z.T.~Zou,  \prd {\bf90}, 114030 (2014).
\bibitem{epjc75-293}
Z.~Rui, W.F.~Wang, G.X.~Wang, L.H.~Song, C.D.~L\"{u}, \epjc {\bf75}, 293 (2015).

\bibitem{epjc77-610}
Z.~Rui, Y.~Li, Z.J.~Xiao,  \epjc {\bf77}, 610 (2017).
\bibitem{epjc60-107}
J.F.~Sun, D.S.~Du, Y.L.~Yang, \epjc {\bf 60}, 107 (2009).
\bibitem{plb763-29}
W.F.~Wang, H.N.~Li,   \plb  {\bf 763}, 29 (2016).

\bibitem{prd95-056008}
Y.~Li, A.J.~Ma, W.F.~Wang, Z.J.~Xiao, Phys. Rev. D {\bf 95}, 056008 (2017).

\bibitem{prd96-093011}
A.J.~Ma, Y.~Li, W.F.~Wang, Z.J.~Xiao, \prd {\bf 96}, 093011 (2017).

\bibitem{prd97-033006}
Z.~Rui, W.F.~Wang,  \prd {\bf 97}, 033006 (2018).

\bibitem{prd98-056019}
Y.~Li, A.J.~Ma, Z.~Rui, W.F.~Wang, Z.J.~Xiao, \prd {\bf 98}, 056019 (2018).

\bibitem{prd97-034033}
C.~Wang, J.B.~Liu, H.N.~Li, C.D.~L\"u,  \prd {\bf 97}, 034033 (2018).

\bibitem{epjc79-37}
Y.~Li, W.F.~Wang, A.J.~Ma, Z.J.~Xiao, \epjc {\bf 79}, 37 (2019).

\bibitem{plb791-342}
W.F.~Wang, J.~Chai, \plb {\bf 791}, 342-350 (2019).

\bibitem{epjc79-539}
A.J.~Ma, W.F.~Wang, Y.~Li, Z.J.~Xiao, \epjc {\bf 79}, 539 (2019).

\bibitem{plb561-258}
C.H.~Chen, H.N.~Li,  Phys. Lett. B  {\bf 561}, 258 (2003).

\bibitem{prd70-054006}
C.H.~Chen, H.N.~Li, \prd  {\bf 70}, 054006 (2004).
\bibitem{prd98-113003}
Z.~Rui, Y.~Li, H.N.~Li, \prd {\bf 98}, 113003 (2018).
\bibitem{1907-04128}
Z.~Rui, Y.~Li, H.~Li, \epjc {\bf 79}, 792 (2019).
\bibitem{1609-07430}
J.~Virto, PoS FPCP {\bf 2016}, 007 (2017).
\bibitem{MP}
D.~M\"uller, D.~Robaschik, B.~Geyer, F.-M.~Dittes, J.~Ho\v rej\v si,  Fortschr. Physik. {\bf 42}, 101 (1994).
\bibitem{MT01}
M.~Diehl, T.~Gousset, B.~Pire, O.~Teryaev, \prl {\bf 81}, 1782 (1998).
\bibitem{MT02}
M.~Diehl, T.~Gousset, B.~Pire, \prd  {\bf 62}, 073014 (2000).
\bibitem{MT03}
Ph.~H\"agler, B.~Pire, L.~Szymanowski, O.V.~Teryaev,  \epjc {\bf 26}, 261 (2002).
\bibitem{NPB555-231}
M.V.~Polyakov,  \npb  {\bf 555},  231 (1999).
\bibitem{Grozin01}
A.G.~Grozin,  Sov. J. Nucl. Phys.  {\bf 38},  289-292 (1983).
\bibitem{Grozin02}
A.G.~Grozin,  Theor. Math. Phys.   {\bf 69}, 1109-1121 (1986).

\bibitem{plb730-336}
U.G.~Mei{\ss}ner, W.~Wang,  \plb  {\bf 730}, 336 (2014).

\bibitem{prd83-112010}
J.P.~Lees et al., [BABAR Collaboration], \prd {\bf 83}, 112010 (2011).


\bibitem{prd70-034009}
Y.~Li,  C.D.~L\"u, Z.J.~Xiao,  and X.Q. ~Yu, \prd {\bf 70},  034009 (2004).

\bibitem{prd76-074018}
A.~Ali, G.~Kramer, Y.~Li, C.D.~L\"u, Y.L.~Shen, W.~Wang, and Y.M.~Wang, \prd {\bf 76},  074018 (2007).

\bibitem{prd85-094003}
Z.J. Xiao, W.F. Wang, and Y.Y. Fan, \prd {\bf 85}, 094003 (2012).

\bibitem{pdg2018}
M.~Tanabashi et al., [Particle Data Group], \prd  {\bf 98}, 030001 (2018).

\bibitem{epjc78-1019}
R.~Aaij et al., [LHCb Collaboration], \epjc {\bf 78}, 1019 (2018).
\bibitem{prd63-054008}
Y.Y.~Keum, H.N.~Li, A.I.~Sanda,  \prd {\bf 63}, 054008 (2001).
\bibitem{epjc23-275}
C.D.~L\"u, M.Z.~Yang,  \epjc {\bf 23}, 275-287 (2002).


\bibitem{prd65-094023}
H.Y.~Cheng, Y.Y.~Keum, K.C.~Yang, \prd {\bf 65}, 094023 (2002).

\bibitem{1909-12524}
Y.~Amhis et al.,   Heavy Flavor Averaging Group (HFLAV), arXiv: 1909.12524 [hep-ex].


\bibitem{zpc1-269}
A.~Ali, J.G.~K\"orner, G.~Kramer, J.~Willrodt, Z. Phys. C {\bf 1}, 269 (1979).

\bibitem{prd64-117503}
M.~Suzuki, Phys. Rev. D {\bf 64}, 117503 (2001).


\end{thebibliography}
\end{document}